\documentclass[a4paper, 10 pt, conference]{ieeeconf}
\IEEEoverridecommandlockouts
\usepackage{amsfonts,amsmath,amssymb} 

\usepackage{psfrag,color}
\usepackage{enumerate,cite,latexsym,graphicx}
\newtheorem{theorem}{Theorem}
\newtheorem{lemma}{Lemma}

\newtheorem{definition}{Definition}

\def\tr{\mathop{\rm Tr}\nolimits} 

\title{\LARGE \bf Robust Stability Analysis of an Optical Parametric Amplifier Quantum System
}

\author{Ian R.~Petersen %
\thanks{This work was supported by the
Australian Research Council (ARC) and Air Force Office of Scientific
Research (AFOSR). This material is based on research sponsored by the
Air Force Research Laboratory, under agreement number
FA2386-09-1-4089.  The U.S. Government is authorized to reproduce and
distribute reprints for Governmental purposes notwithstanding any
copyright notation thereon.
The views and conclusions contained herein are those of the authors
and should not be interpreted as necessarily representing the official
policies or endorsements, either expressed or implied, of the Air
Force Research Laboratory or the U.S. Government. }%
\thanks{Ian R. Petersen is with the School of  Engineering and Information Technology, 
        University of New South Wales at the Australian Defence Force Academy, Canberra ACT 2600, Australia.
         {\tt\small i.r.petersen@gmail.com} } 
}%

\begin{document}

\maketitle
\thispagestyle{empty}
\pagestyle{empty}

\begin{abstract}
This paper considers the problem of robust stability for a class of
uncertain nonlinear quantum systems subject to unknown perturbations in the
system Hamiltonian.  The
 case of a nominal linear quantum system is considered with
non-quadratic perturbations to the system
Hamiltonian. 
The paper extends recent results on the robust stability of nonlinear quantum systems to allow for non-quadratic perturbations to the Hamiltonian which depend on multiple parameters. A robust stability condition is given in terms of a strict bounded real condition. This result is then applied to the robust stability analysis of a nonlinear quantum system which is a model of an optical parametric amplifier. 
\end{abstract}

\section{Introduction} \label{sec:intro}
In recent years, a number of papers have considered the feedback
control of systems whose dynamics are governed by the laws of quantum
mechanics rather than classical mechanics; e.g., see
\cite{YK03A,YK03B,YAM06,JNP1,NJP1,GGY08,MaP3,MaP4,YNJP1,GJ09,GJN10,WM10,PET10Ba}. In
particular, the papers \cite{GJ09,JG10} consider a framework of
quantum systems defined in terms of a triple $(S,L,H)$ where $S$ is a
scattering matrix, $L$ is a vector of coupling operators and $H$ is a
Hamiltonian operator. 

The papers \cite{PUJ1a,PUJ2} consider the problem of absolute stability for a quantum system defined in terms of a triple $(S,L,H)$ in which the quantum system Hamiltonian is
decomposed as $H =H_1+H_2$ where $H_1$ is a known nominal Hamiltonian
and $H_2$ is a perturbation Hamiltonian, which is contained in a
specified set of Hamiltonians $\mathcal{W}$. In particular the papers \cite{PUJ1a,PUJ2} consider the case in which the
nominal Hamiltonian $H_1$ is a quadratic function of annihilation and
creation operators and the coupling operator vector is a linear
function of annihilation and creation operators. This case corresponds
to a nominal  linear quantum system; e.g., see
\cite{JNP1,NJP1,MaP3,MaP4,PET10Ba}. The results in \cite{PUJ1a,PUJ2} have recently been extended to allow for uncertainties in the coupling operator $L$ \cite{PUJ3a}. Also, the results of \cite{PUJ1a} have been applied to the robust stability analysis of  a quantum system
which consists of a Josephson junction in a resonant cavity \cite{PET12Aa}. 

In the paper  \cite{PUJ1a}, it is assumed that $H_2$ is contained in a set of non-quadratic perturbation Hamiltonians corresponding to a sector bound on the nonlinearity. In this case, \cite{PUJ1a} obtains a robust stability result in terms of a frequency domain condition. Also, the paper \cite{PUJ2} restricts attention to quadratic perturbation Hamiltonians. In this case, which corresponds to linear perturbed quantum systems, a frequency domain robust stability condition is also obtained. 

An example considered in the paper \cite{PUJ2} involves the robust stability analysis of a quantum system consisting of a linearized optical parametric amplifier (OPA). Optical parametric amplifiers are widely used in the field of experimental quantum optics; e.g., see \cite{GZ00,BR04,WM08}. In particular, they can be used as optical squeezers which produce squeezed light which has  a smaller  noise variance in one quadrature than the standard quantum limit. This is at the expense of a larger noise variance in the other quadrature; e.g., see \cite{GZ00,BR04,WM08,SHHPJ2a,SaP2a,SaPH1a}. Such an OPA can be produced by 
enclosing a second-order nonlinear optical medium in an optical cavity; e.g., see \cite{BR04,SHHPJ2a,SaP2a,SaPH1a}. Thus, an OPA is an inherently nonlinear quantum system. However, the paper \cite{PUJ2} only dealt with linear perturbed quantum systems and thus the results of this paper could only be used to analyze the robust stability of a linearized version of the OPA. Furthermore the results of \cite{PUJ1a} on nonlinear perturbed quantum systems cannot be directly applied to the OPA system since the results of \cite{PUJ1a} only deal with scalar nonlinearities but the nonlinearity in the OPA model is dependent on two variables; e.g., see \cite{BR04,WM08}. 
In this paper, we extend  the result of \cite{PUJ1a} on the robust stability of nonlinear quantum systems to allow for non-quadratic perturbations in the Hamiltonian which depend on multiple variables. This enables us to analyze the
robust stability of the OPA nonlinear quantum system. 

\section{Robust Stability of Uncertain Nonlinear Quantum Systems} \label{sec:systems}
In this section, we describe the general class of quantum systems under consideration. 
As in the papers \cite{GJ09,JG10,PUJ1a,PUJ2,JPU1a},  we consider uncertain nonlinear open quantum systems defined by  parameters $(S,L,H)$ where $S$ is the scattering matrix which is typically chosen as the identity matrix, L is the coupling operator and $H$ is the system  Hamiltonian operator which is assumed to be of the form
\begin{equation}
\label{H1}
H = \frac{1}{2}\left[\begin{array}{cc}a^\dagger &
      a^T\end{array}\right]M
\left[\begin{array}{c}a \\ a^\#\end{array}\right]+f(z,z^*).
\end{equation}
Here $a$ is a vector of annihilation
operators on the underlying Hilbert space and $a^\#$ is the
corresponding vector of creation operators. Also,  $M \in \mathbb{C}^{2n\times 2n}$ is a Hermitian matrix of the
form
\begin{equation}
\label{Mform}
M= \left[\begin{array}{cc}M_1 & M_2\\
M_2^\# &     M_1^\#\end{array}\right]
\end{equation}
and $M_1 = M_1^\dagger$, $M_2 = M_2^T$.
In the case vectors of
operators, the notation  $^\dagger$ refers to the transpose of the vector of adjoint
operators and  in the case of matrices, this notation refers to the complex conjugate transpose of a matrix. In the case vectors of
operators, the notation $^\#$ refers to the vector of adjoint
operators and in the case of complex matrices, this notation refers to
the complex conjugate matrix. Also, the notation $^*$ denotes the adjoint of an
operator. The matrix $M$ is assumed to be known and defines the nominal quadratic part of the system Hamiltonian. 
  Furthermore, we assume the uncertain non-quadratic  part of the system Hamiltonian  $f(z,z^*)$ is defined by a formal power series of  the form
\begin{eqnarray}
\label{H2nonquad}
f(z,z^*) &=& \sum_{i=1}^p\sum_{j=1}^p\sum_{k=0}^\infty\sum_{\ell=0}^\infty S_{ijk\ell}z_i^k(z_j^*)^\ell \nonumber \\
&=& \sum_{i=1}^p\sum_{j=1}^p\sum_{k=0}^\infty \sum_{\ell=0}^\infty S_{ijk\ell} H_{ijk\ell}
\end{eqnarray}
which is assumed to converge in some suitable sense.
Here $S_{ijk\ell}=S_{ji\ell k}^*$, $H_{ijk\ell} = z_i^k(z_j^*)^\ell$, and $z = \left[\begin{array}{llll} z_1 & z_2 & \ldots & z_m \end{array}\right]^T$ is a vector of  operators on the underlying Hilbert space defined by
\begin{eqnarray}
\label{z}
z &=&  E_1a+E_2 a^\# \nonumber \\
&=& \left[\begin{array}{cc} E_1 & E_2 \end{array}\right]
\left[\begin{array}{c}a \\ a^\#\end{array}\right] = \tilde E 
\left[\begin{array}{c}a \\ a^\#\end{array}\right].
\end{eqnarray}
Also, we write 
\[
\tilde E  = \left[\begin{array}{c}\tilde E_1 \\ \tilde E_2\\ \vdots \\ \tilde E_p \end{array}\right].
\]
It follows from this definition of $f(z,z^*)$ that
\begin{eqnarray*}
f(z,z^*)^* &=& \sum_{i=1}^p\sum_{j=1}^p\sum_{k=0}^\infty\sum_{\ell=0}^\infty S_{ijk\ell}^*z_j^\ell(z_i^*)^k \\
&=& 
\sum_{j=1}^p\sum_{i=1}^p\sum_{\ell=0}^\infty\sum_{k=0}^\infty S_{ji\ell k}z_j^\ell(z_i^*)^k = f(z,z^*)
\end{eqnarray*}
and thus $f(z,z^*)$ is a self-adjoint operator. 
The term $f(z,z^*)$ is referred to as the perturbation Hamiltonian. It  is assumed to be unknown but is contained within a known set which will be defined below. 

We assume the coupling operator $L$ is known and is of the form 
\begin{equation}
\label{L}
L = \left[\begin{array}{cc}N_1 & N_2 \end{array}\right]
\left[\begin{array}{c}a \\ a^\#\end{array}\right]
\end{equation}
where $N_1 \in \mathbb{C}^{m\times n}$ and $N_2 \in
\mathbb{C}^{m\times n}$. Also, we write
\[
\left[\begin{array}{c}L \\ L^\#\end{array}\right] = N
\left[\begin{array}{c}a \\ a^\#\end{array}\right] =
\left[\begin{array}{cc}N_1 & N_2\\
N_2^\# &     N_1^\#\end{array}\right]
\left[\begin{array}{c}a \\ a^\#\end{array}\right].
\]  

The annihilation and creation operators are assumed to satisfy the
canonical commutation relations:
\begin{eqnarray}
\label{CCR2}
\left[\left[\begin{array}{l}
      a\\a^\#\end{array}\right],\left[\begin{array}{l}
      a\\a^\#\end{array}\right]^\dagger\right]
&\stackrel{\Delta}{=}&\left[\begin{array}{l} a\\a^\#\end{array}\right]
\left[\begin{array}{l} a\\a^\#\end{array}\right]^\dagger
\nonumber \\
&&- \left(\left[\begin{array}{l} a\\a^\#\end{array}\right]^\#
\left[\begin{array}{l} a\\a^\#\end{array}\right]^T\right)^T\nonumber \\
&=& J
\end{eqnarray}
where $J = \left[\begin{array}{cc}I & 0\\
0 & -I\end{array}\right]$; e.g., see \cite{GGY08,GJN10,PET10Ba}.

To define the set of allowable perturbation Hamiltonians $f(\cdot)$, we first define the following formal partial derivatives:
\begin{equation}
\label{fdash}
\frac{\partial f(z,z^*)}{\partial z_i} = \sum_{j=1}^p\sum_{k=1}^\infty\sum_{\ell=0}^\infty kS_{ijk\ell}z_i^{k-1}(z_j^*)^\ell,
\end{equation}
\begin{equation}
\label{fddash}
\frac{\partial^2 f(z,z^*)}{\partial z_i^2} 
= \sum_{j=1}^p\sum_{k=1}^\infty\sum_{\ell=0}^\infty k(k-1)S_{ijk\ell}z_i^{k-2}(z_j^*)^\ell
\end{equation}
and for given constants $\gamma > 0$,  $\delta_1\geq 0$, $\delta_2\geq 0$, we consider the 
sector bound condition
\begin{equation}
\label{sector4a}
\sum_{i=1}^p\frac{\partial f(z,z^*)}{\partial z_i}^*\frac{\partial f(z,z^*)}{\partial z_i}  
\leq \frac{1}{\gamma^2}\sum_{i=1}^pz_i z_i^* + \delta_1
\end{equation}
and the condition
\begin{equation}
\label{sector4b}
\sum_{i=1}^p\frac{\partial^2 f(z,z^*)}{\partial z_i^2}^*\frac{\partial^2 f(z,z^*)}{\partial z_i^2} \leq  \delta_2.
\end{equation}
Then we define the set of perturbation Hamiltonians $\mathcal{W}$  as follows:
\begin{equation}
\label{W5}
\mathcal{W} = \left\{\begin{array}{l}f(\cdot) \mbox{ of the form
      (\ref{H2nonquad}) such that 
} \\
\mbox{ conditions (\ref{sector4a}) and (\ref{sector4b}) are satisfied}\end{array}\right\}.
\end{equation}
Note that the condition (\ref{sector4b}) effectively amounts to a global Lipschitz condition on the quantum nonlinearity. 

As in \cite{PUJ1a,PUJ2}, we will consider the following notion of robust mean square stability. 
\begin{definition}
\label{D1}
An uncertain open quantum system defined by  $(S,L,H)$ where $H$ is of the form (\ref{H1}), $f(\cdot) \in \mathcal{W}$, and $L$  of the form (\ref{L}) is said to be {\em robustly mean square stable} if there exist constants $c_1 > 0$, $c_2 > 0$ and $c_3 \geq 0$ such that for any $f(\cdot) \in \mathcal{W}$,  
\begin{eqnarray}
\label{ms_stable0}
\lefteqn{\left< \left[\begin{array}{c}a(t) \\ a^\#(t)\end{array}\right]^\dagger \left[\begin{array}{c}a(t) \\ a^\#(t)\end{array}\right] \right>}\nonumber \\
&\leq& c_1e^{-c_2t}\left< \left[\begin{array}{c}a \\ a^\#\end{array}\right]^\dagger \left[\begin{array}{c}a \\ a^\#\end{array}\right] \right>
+ c_3~~\forall t \geq 0.
\end{eqnarray}
Here $\left[\begin{array}{c}a(t) \\ a^\#(t)\end{array}\right]$ denotes the Heisenberg evolution of the vector of operators $\left[\begin{array}{c}a \\ a^\#\end{array}\right]$; e.g., see \cite{JG10}.
\end{definition}

We will show that  the following small gain condition
is sufficient for the robust mean square stability
of the nonlinear quantum system under consideration when $f(\cdot) \in \mathcal{W}$: 
\begin{enumerate}
\item
The matrix 
\begin{equation}
\label{Hurwitz1}
F = -\imath JM-\frac{1}{2}JN^\dagger J N\mbox{ is Hurwitz;}
\end{equation}
\item
\begin{equation}
\label{Hinfbound1}
\left\|\tilde E^\# \Sigma\left(sI -F\right)^{-1}J\Sigma \tilde E^T \right\|_\infty < \frac{\gamma}{2}
\end{equation}
where  $\Sigma = \left[\begin{array}{cc} 0 & I\\
I &0 \end{array}\right].
$
\end{enumerate}
This leads to the following theorem.

\begin{theorem}
\label{T4}
Consider an uncertain open nonlinear quantum system defined by $(S,L,H)$  such that
$H$ is of the form (\ref{H1}), $L$ is of the
form (\ref{L}) and $f(\cdot) \in \mathcal{W}$. Furthermore, assume that
the strict bounded real condition  (\ref{Hurwitz1}), (\ref{Hinfbound1})
is satisfied. Then the
uncertain quantum system is robustly mean square stable.
\end{theorem}
In order to prove this theorem, we require the following definitions and lemmas.

\begin{lemma}[See Lemma 3.4 of \cite{JG10}.]
\label{L0}
Consider an open quantum system defined by $(S,L,H)$ and suppose there exists a non-negative self-adjoint operator $V$ on the underlying Hilbert space such that
\begin{equation}
\label{lyap_ineq}
-\imath[V,H] + \frac{1}{2}L^\dagger[V,L]+\frac{1}{2}[L^\dagger,V]L + cV \leq \lambda
\end{equation}
where $c > 0$ and $\lambda$ are real numbers. 
Then for any plant state, we have
\[
\left<V(t)\right> \leq e^{-ct}\left<V\right> + \frac{\lambda}{c},~~\forall t \geq 0.
\]
\end{lemma}

In the above lemma, $[\cdot,\cdot]$ denotes the commutator between two operators. In the case of a commutator between a scalar operator and a vector of operators, this notation denotes the corresponding vector of commutator operators. Also, $V(t)$ denotes the Heisenberg evolution of the operator $V$ and $\left<\cdot\right>$ denotes quantum expectation; e.g., see \cite{JG10}.

We will consider quadratic ``Lyapunov'' operators  $V$ of the form 
\begin{equation}
\label{quadV}
V = \left[\begin{array}{cc}a^\dagger &
      a^T\end{array}\right]P
\left[\begin{array}{c}a \\ a^\#\end{array}\right]
\end{equation}
where $P \in \mathbb{C}^{2n\times 2n}$ is a positive-definite Hermitian matrix of the
form
\begin{equation}
\label{Pform}
P= \left[\begin{array}{cc}P_1 & P_2\\
P_2^\# &     P_1^\#\end{array}\right].
\end{equation}
 Hence, we consider a set of  non-negative self-adjoint operators
$\mathcal{P}$ defined as
\begin{equation}
\label{P1}
\mathcal{P} = \left\{\begin{array}{l}V \mbox{ of the form
      (\ref{quadV}) such that $P > 0$ is a 
} \\
\mbox{  Hermitian matrix of the form (\ref{Pform})}\end{array}\right\}.
\end{equation}

\begin{lemma}
\label{L4}
Given any $V \in \mathcal{P}$, then
\begin{equation}
\label{mui}
\mu_i = \left[z_i,[z_i,V]\right] = \left[z_i^*,[z_i^*,V]\right]^* = 
-\tilde E_i \Sigma JPJ\tilde E_i^T,
\end{equation}
which are constants for $i=1,2,\ldots,p$.  
\end{lemma}
{\em Proof:}
The proof of this result follows via a straightforward but tedious
calculation using (\ref{CCR2}). \hfill $\Box$

\begin{lemma}
\label{LB}
Given any $V \in \mathcal{P}$, then
\begin{eqnarray}
\label{comm_condition}
[V,f(z,z^*)] &=& \sum_{i=1}^p[V,z_i]w_{1i}^* -\sum_{i=1}^pw_{1i}[z_i^*,V]\nonumber \\
&&+ \frac{1}{2}\sum_{i=1}^p\mu_i w_{2i}^*\nonumber \\
&&-\frac{1}{2}\sum_{i=1}^p w_{2i}\mu_i^*
\end{eqnarray} 
where
 $z=\left[\begin{array}{llll} z_1 & z_2 & \ldots & z_p \end{array}\right]^T$,
\begin{eqnarray}
\label{zw1w2}
w_1&=& \left[\begin{array}{llll} w_{11} & w_{12} & \ldots & w_{1p} \end{array}\right]^T,~
w_{1i} = \frac{\partial f(z,z^*)}{\partial z_i}^*,\nonumber \\
w_2&=& \left[\begin{array}{llll} w_{21} & w_{22} & \ldots & w_{2p} \end{array}\right]^T,~ w_{2i} = \frac{\partial^2 f(z,z^*)}{\partial z_i^2}^*,\nonumber \\
\end{eqnarray}
and the constants $\mu_i$ are defined as in (\ref{mui}). 
\end{lemma}

\noindent
{\em Proof:}
First, we note that given any $V \in \mathcal{P}$, $i \in \left\{1,2,\ldots,p\right\}$, and $k \geq 1$,
\begin{eqnarray}
\label{Vzetak}
 Vz_i  &=& [V,z_i]+ z_i V;\nonumber \\
\vdots && \nonumber \\
Vz_i^k &=& \sum_{n=1}^k z_i^{n-1}[V,z_i] z_i^{k-n}+z_i^k V.
\end{eqnarray}
Also using Lemma \ref{L4}, it follows that for any $n \geq 1$,
\begin{eqnarray}
\label{Vzetak1}
z_i [V,z_i] &=& [V,z_i]z_i + \mu_i; \nonumber \\
\vdots && \nonumber \\
 z_i^{n-1} [V,z_i]&=&[V,z_i]z_i^{n-1}
+ (n-1)z_i^{n-2}\mu_i.
\end{eqnarray}
Therefore using (\ref{Vzetak}) and (\ref{Vzetak1}), it follows that
\begin{eqnarray*}
Vz_i^k &=&\sum_{n=1}^k [V,z_i] z_i^{n-1}z_i^{k-n}+ (n-1)z_i^{n-2}z_i^{k-n}\mu_i  \nonumber \\
&&+z_i^k V\nonumber \\
&=& \sum_{n=1}^k [V,z_i] z_i^{k-1}+ (n-1)z_i^{k-2}\mu_i  +z_i^k V\nonumber \\
&=&k[V,z_i] z_i^{k-1}+\frac{k(k-1)}{2}z_i^{k-2}\mu_i+z_i^k V 
\end{eqnarray*}
which holds for any $i \in \left\{1,2,\ldots,p\right\}$ and $k \geq 0$. 
Similarly for any $j \in \left\{1,2,\ldots,p\right\}$ and $\ell \geq 0$,
\begin{eqnarray*}
(z_j^*)^\ell V &=& \ell(z_j^*)^{\ell-1}[z_j^*,V]+\frac{\ell(\ell-1)}{2}\mu_j^*(z_j^*)^{\ell-2}
\nonumber \\
&&+V(z_j^*)^\ell.
\end{eqnarray*}

Now given any $i \in \left\{1,2,\ldots,p\right\}$, $j \in \left\{1,2,\ldots,p\right\}$, $k \geq 0$, $\ell \geq 0$,  we have using the notation in (\ref{H2nonquad}):
\begin{eqnarray}
\label{VHkl}
\lefteqn{[V,H_{ijk\ell}]}\nonumber \\
 &=& k[V,z_i] z_i^{k-1}(z_j^*)^\ell+\frac{k(k-1)}{2}\mu_iz_i^{k-2}(z_j^*)^\ell\nonumber \\&&
+z_i^k V(z_j^*)^\ell\nonumber \\
&& -\ell z_i^k(z_j^*)^{\ell-1}[z_j^*,V]-\frac{\ell(\ell-1)}{2}\mu_j^*z_i^k(z_j^*)^{\ell-2}\nonumber \\&&
-z_i^k V(z_j^*)^\ell\nonumber \\
&=& k[V,z_i] z_i^{k-1}(z_j^*)^\ell-\ell z_i^k(z_j^*)^{\ell-1}[z_j^*,V]\nonumber \\&&
+\frac{k(k-1)}{2}\mu_iz_i^{k-2}(z_j^*)^\ell-\frac{\ell(\ell-1)}{2}\mu_j^*z_i^k(z_j^*)^{\ell-2}.\nonumber \\
\end{eqnarray}
Therefore,
\begin{eqnarray}
\label{VH2}
[V,f(z,z^*)] &=& \sum_{i=1}^p\sum_{j=1}^p\sum_{k=0}^\infty \sum_{\ell=0}^\infty S_{ijk\ell} [V,H_{ijk\ell}] \nonumber \\
&=& \sum_{i=1}^p[V,z_i]\frac{\partial f(z,z^*)}{\partial z_i} \nonumber \\
&&-\sum_{j=1}^p\frac{\partial f(z,z^*)}{\partial z_j}^*[z_j^*,V]\nonumber \\
&&+ \frac{1}{2}\sum_{i=1}^p\mu_i \frac{\partial^2 f(z,z^*)}{\partial z_i^2}\nonumber \\
&&-\frac{1}{2}\sum_{j=1}^p \frac{\partial^2 f(z,z^*)}{\partial z_j^2}^*\mu_j^*.
\end{eqnarray}
Now 
 it follows from (\ref{zw1w2}) that condition (\ref{comm_condition}) is
 satisfied. 
 respectively.
\hfill $\Box$

\begin{lemma}
\label{L2}
Given $V \in \mathcal{P}$ and $L$ defined as in (\ref{L}), then
\begin{eqnarray*}
\lefteqn{[V,\frac{1}{2}\left[\begin{array}{cc}a^\dagger &
      a^T\end{array}\right]M
\left[\begin{array}{c}a \\ a^\#\end{array}\right]] =}\nonumber \\
&& \left[\left[\begin{array}{cc}a^\dagger &
      a^T\end{array}\right]P
\left[\begin{array}{c}a \\ a^\#\end{array}\right],\frac{1}{2}\left[\begin{array}{cc}a^\dagger &
      a^T\end{array}\right]M
\left[\begin{array}{c}a \\ a^\#\end{array}\right]\right] \nonumber \\
&=& \left[\begin{array}{c}a \\ a^\#\end{array}\right]^\dagger 
\left[
PJM - MJP 
\right] \left[\begin{array}{c}a \\ a^\#\end{array}\right].
\end{eqnarray*}
Also,
\begin{eqnarray*}
\lefteqn{\frac{1}{2}L^\dagger[V,L]+\frac{1}{2}[L^\dagger,V]L =} \nonumber \\
&=& \tr\left(PJN^\dagger\left[\begin{array}{cc}I & 0 \\ 0 & 0 \end{array}\right]NJ\right)
\nonumber \\&&
-\frac{1}{2}\left[\begin{array}{c}a \\ a^\#\end{array}\right]^\dagger
\left(N^\dagger J N JP+PJN^\dagger J N\right)
\left[\begin{array}{c}a \\ a^\#\end{array}\right].
\end{eqnarray*}
Furthermore, 
\[
\left[\left[\begin{array}{c}a \\ a^\#\end{array}\right],\left[\begin{array}{cc}a^\dagger &
      a^T\end{array}\right]P
\left[\begin{array}{c}a \\ a^\#\end{array}\right]\right] = 2JP\left[\begin{array}{c}a \\ a^\#\end{array}\right].
\]
\end{lemma}
{\em Proof:}
The proof of these identities follows via  straightforward but tedious
calculations using (\ref{CCR2}). \hfill $\Box$

\noindent
{\em Proof of Theorem \ref{T4}.}
It follows from (\ref{z}) that we can write
\begin{eqnarray*}
z^\# &=& E_1^\#a^\#+E_2^\# a=\left[\begin{array}{cc} E_2^\# & E_1^\# \end{array}\right]
\left[\begin{array}{c}a \\ a^\#\end{array}\right]\nonumber \\
&=&  \tilde E^\# \Sigma \left[\begin{array}{c}a \\ a^\#\end{array}\right].
\end{eqnarray*}
Also,  it follows from Lemma \ref{L2} that
\[
[z^\#,V] = 2 \tilde E^\# \Sigma
JP\left[\begin{array}{c}a \\ a^\#\end{array}\right].
\]
Furthermore, $[V,z^T] = [z^\#,V]^\dagger$ and  hence,
\begin{eqnarray}
\label{VzzV}
[V,z^T] [z^\#,V] =
4\left[\begin{array}{c}a \\ a^\#\end{array}\right]^\dagger PJ 
\Sigma \tilde E^T \tilde E^\# \Sigma
JP
\left[\begin{array}{c}a \\ a^\#\end{array}\right].
\end{eqnarray}
Also, we can write
\begin{equation}
\label{zz}
z^Tz^\# = \left[\begin{array}{c}a \\ a^\#\end{array}\right]^\dagger
\Sigma \tilde E^T \tilde E^\# \Sigma
\left[\begin{array}{c}a \\ a^\#\end{array}\right].
\end{equation}

Hence using Lemma \ref{L2}, we obtain
\begin{eqnarray}
\label{lyap_ineq3}
&&-\imath[V,\frac{1}{2}\left[\begin{array}{cc}a^\dagger &
      a^T\end{array}\right]M
\left[\begin{array}{c}a \\ a^\#\end{array}\right]]\nonumber \\
&&+ \frac{1}{2}L^\dagger[V,L]+\frac{1}{2}[L^\dagger,V]L
+ [V,z^T][z^\#,V]
+\frac{z^Tz^\#}{\gamma^2}
 \nonumber \\
&=& \left[\begin{array}{c}a \\ a^\#\end{array}\right]^\dagger\left(\begin{array}{c}
F^\dagger P + P F\\ 
+4 PJ\Sigma \tilde E^T \tilde E^\# \Sigma JP \\
+ \frac{1}{ \gamma^2}\Sigma \tilde E^T \tilde E^\# \Sigma\\
\end{array}\right)\left[\begin{array}{c}a \\
a^\#\end{array}\right]\nonumber \\
&&+\tr\left(PJN^\dagger\left[\begin{array}{cc}I & 0 \\ 0 & 0 \end{array}\right]NJ\right)
\end{eqnarray}
where $F = -\imath JM-\frac{1}{2}JN^\dagger J N$. 

We now observe that using the  strict bounded real lemma, (\ref{Hurwitz1}) and (\ref{Hinfbound1}) imply  that the matrix inequality 
\begin{equation}
\label{QMI2}
F^\dagger P + P F 
+4 PJ\Sigma \tilde E^T \tilde E^\# \Sigma JP 
+ \frac{1}{ \gamma^2}\Sigma \tilde E^T \tilde E^\# \Sigma
 < 0.
\end{equation}
will have a solution $P > 0$ of the form (\ref{Pform}); e.g., see \cite{ZDG96,MaP4}.  This matrix $P$ defines a corresponding operator $V \in \mathcal{P}_1$ as in (\ref{quadV}). From this, it follows using (\ref{lyap_ineq3}) that there exists a constant $c > 0$ such that 
\begin{eqnarray}
\label{dissip1a}
&&-\imath[V,\frac{1}{2}\left[\begin{array}{cc}a^\dagger &
      a^T\end{array}\right]M
\left[\begin{array}{c}a \\ a^\#\end{array}\right]]\nonumber \\
&&+ \frac{1}{2}L^\dagger[V,L]+\frac{1}{2}[L^\dagger,V]L
+ \sum_{i=1}^p[V,z_i][z_i^*,V]
\nonumber \\&&
+ \frac{1}{ \gamma^2}\sum_{i=1}^p z_iz_i^*
+ cV \leq \tilde \lambda.
\nonumber \\
\end{eqnarray}
with 
\[
\tilde \lambda = \tr\left(PJN^\dagger\left[\begin{array}{cc}I & 0 \\ 0 & 0 \end{array}\right]NJ\right) \geq 0.
\]
Also, it follows from Lemma \ref{LB} that
\begin{eqnarray}
\label{ineq1a}
\lefteqn{-\imath[V,H] + \frac{1}{2}L^\dagger[V,L]+\frac{1}{2}[L^\dagger,V]L}\nonumber \\
 &=& -\imath[V,f(z,z^*)]-\imath[V,\frac{1}{2}\left[\begin{array}{cc}a^\dagger &
      a^T\end{array}\right]M
\left[\begin{array}{c}a \\ a^\#\end{array}\right]]\nonumber \\
&&+ \frac{1}{2}L^\dagger[V,L]+\frac{1}{2}[L^\dagger,V]L\nonumber \\
&=&-\imath[V,\frac{1}{2}\left[\begin{array}{cc}a^\dagger &
      a^T\end{array}\right]M
\left[\begin{array}{c}a \\ a^\#\end{array}\right]]\nonumber \\
&&+ \frac{1}{2}L^\dagger[V,L]+\frac{1}{2}[L^\dagger,V]L\nonumber \\
&&-\imath\sum_{i=1}^p[V,z_i]w_{1i}^*+\imath\sum_{i=1}^pw_{1i}[z_i^*,V]\nonumber \\
&&-\frac{1}{2}\imath\sum_{i=1}^p\mu_i w_{2i}^*+\frac{1}{2}\imath \sum_{i=1}^p w_{2i}\mu_i^*. 
\end{eqnarray}
Furthermore, $[V,z_i]^* = z_i^*V-Vz_i^*=[z_i^*,V]$ since $V$ is self-adjoint. Therefore, 
\begin{eqnarray*}
0 &\leq& \sum_{i=1}^p\left([V,z_i]- \imath w_{1i}\right)
\left([V,z_i]- \imath w_{1i}\right)^*\nonumber \\
&=&  \sum_{i=1}^p[V,z_i][z_i^*,V]+\imath\sum_{i=1}^p[V,z_i]w_{1i}^*\nonumber \\
&&-\imath \sum_{i=1}^p w_{1i}[z_i^*,V]+\sum_{i=1}^p w_{1i} w_{1i}^*
\end{eqnarray*}
and hence
\begin{eqnarray}
\label{ineq3a}
 \lefteqn{-\imath\sum_{i=1}^p[V,z_i]w_{1i}^*+\imath\sum_{i=1}^pw_{1i}[z_i^*,V]}\nonumber \\
&\leq& \sum_{i=1}^p[V,z_i][z_i^*,V]+\sum_{i=1}^p w_{1i} w_{1i}^*.
\end{eqnarray}

Also, 
\begin{eqnarray*}
0 &\leq& \sum_{i=1}^p \left(\frac{1}{2}\mu_i- \imath w_{2i}\right)
\left(\frac{1}{2}\mu_i- \imath w_{2i}\right)^*\nonumber \\
&=&  \frac{1}{4} \sum_{i=1}^p \mu_i \mu_i^*-\frac{\imath}{2} \sum_{i=1}^p w_{2i}\mu_i^*+\frac{\imath}{2}\sum_{i=1}^p \mu_i w_{2i}^*
\nonumber \\
&&+ \sum_{i=1}^p w_{2i}  w_{2i}^*
\end{eqnarray*}
and hence
\begin{eqnarray}
\label{ineq3b}
\lefteqn{\frac{\imath}{2}\sum_{i=1}^p w_{2i}\mu_i^*-\frac{\imath}{2}\sum_{i=1}^p \mu_i w_{2i}^*}\nonumber \\
 &\leq& \frac{1}{4}\sum_{i=1}^p \mu_i\mu_i^*
+\sum_{i=1}^p w_{2i} w_{2i}^*.
\end{eqnarray}
Furthermore, it follows from (\ref{sector4a}) and (\ref{sector4b}) that
\begin{equation}
\label{sector2a}
\sum_{i=1}^p w_{1i} w_{1i}^* \leq \frac{1}{\gamma^2}\sum_{i=1}^p z_i z_i^* + \delta_1
\end{equation}
and
\begin{equation}
\label{sector2b}
\sum_{i=1}^p w_{2i} w_{2i}^* \leq \delta_2.
\end{equation}
Substituting (\ref{ineq3a}), (\ref{ineq3b}), (\ref{sector2a}) and (\ref{sector2b}) into (\ref{ineq1a}), it follows that
\begin{eqnarray}
\label{ineq2a}
\lefteqn{-\imath[V,H] + \frac{1}{2}L^\dagger[V,L]+\frac{1}{2}[L^\dagger,V]L}\nonumber \\
 &\leq &  -\imath[V,\frac{1}{2}\left[\begin{array}{cc}a^\dagger &
      a^T\end{array}\right]M
\left[\begin{array}{c}a \\ a^\#\end{array}\right]]\nonumber \\
&& + \frac{1}{2}L^\dagger[V,L]+\frac{1}{2}[L^\dagger,V]L\nonumber \\
&&+ \sum_{i=1}^p [V,z_i][z_i^*,V]
\nonumber \\&&
+\frac{1}{\gamma^2} \sum_{i=1}^p z_i z_i^* 
+\delta_1
+\sum_{i=1}^p \mu_i\mu_i^*/4+\delta_2
\end{eqnarray}

Then it follows from (\ref{dissip1a}) that 
\begin{eqnarray*}
\lefteqn{-\imath[V,H] + \frac{1}{2}L^\dagger[V,L]+\frac{1}{2}[L^\dagger,V]L + cV }\nonumber \\
&& \leq \tilde \lambda + \delta_1+\sum_{i=1}^p \mu_i\mu_i^*/4+\delta_2. 
\end{eqnarray*}
From this,  it follows from Lemma \ref{L0} and  $P > 0$ that 
\begin{eqnarray}
\label{ms_stable1}
\lefteqn{\left< \left[\begin{array}{c}a(t) \\ a^\#(t)\end{array}\right]^\dagger \left[\begin{array}{c}a(t) \\ a^\#(t)\end{array}\right] \right>}\nonumber \\
&\leq&  e^{-ct}\left< \left[\begin{array}{c}a(0) \\ a^\#(0)\end{array}\right]^\dagger \left[\begin{array}{c}a(0) \\ a^\#(0)\end{array}\right] \right>\frac{\lambda_{max}[P]}{\lambda_{min}[P]}\nonumber \\
&&+ \frac{\lambda}{c\lambda_{min}[P]}~~\forall t \geq 0
\end{eqnarray} 
where $\lambda = \tilde \lambda+ \delta_1+\sum_{i=0}^p\mu_i\mu_i^*/4+\delta_2.$
 Hence, the condition (\ref{ms_stable0}) is satisfied with $c_1 = \frac{\lambda_{max}[P]}{\lambda_{min}[P]} > 0$, $c_2 = c > 0$ and $c_3 = \frac{\lambda}{c\lambda_{min}[P]} \geq 0$. 
\hfill $\Box$

Note that the strict bounded real condition  (\ref{Hinfbound1}) can be simplified according the following lemma.
\begin{lemma}
\label{L3}
The strict bounded real condition  (\ref{Hinfbound1}) is satisfied if and only if the following equivalent strict bounded real condition is satisfied:
\begin{equation}
\label{Hinfbound2}
\left\|\tilde E\left(sI -F\right)^{-1}J\tilde E^\dagger  \right\|_\infty < \frac{\gamma}{2}.
\end{equation}
\end{lemma}

{\em Proof:}
First note that since $\Sigma = \Sigma^{-1}$, we can write
\begin{eqnarray}
\label{TF1}
\tilde E^\# \Sigma\left(sI -F\right)^{-1}\tilde D &=& \tilde E^\# \Sigma\left(sI -F\right)^{-1}J\Sigma \tilde E^T\nonumber \\
&=& \tilde E^\# \left(sI -\Sigma F\Sigma\right)^{-1}\Sigma J\Sigma \tilde E^T.\nonumber \\
\end{eqnarray}
Furthermore, it is straightforward to verify that $\Sigma J\Sigma=-J$, $\Sigma M\Sigma = M^\#$, $\Sigma N\Sigma = N^\#$, 
$\Sigma N^\dagger \Sigma = N^T$ and hence
\begin{eqnarray*}
\Sigma F\Sigma &=& -\imath\Sigma J M \Sigma -\frac{1}{2}\Sigma JN^\dagger J N \Sigma \\
&=& -\imath\Sigma J \Sigma \Sigma M \Sigma -\frac{1}{2}\Sigma J\Sigma \Sigma N^\dagger \Sigma \Sigma J \Sigma \Sigma N \Sigma \\
&=& \imath JM^\# - \frac{1}{2}JN^TJN^\# \\
&=& F^\#.
\end{eqnarray*}
Therefore, it follows from (\ref{TF1}) that
\begin{eqnarray*}
\tilde E^\# \Sigma\left(sI -F\right)^{-1}\tilde D &=& -\tilde E^\# \left(sI - F^\#\right)^{-1}J \tilde E^T \\
&=& -\left(\tilde E\left(s^*I -F\right)^{-1}J\tilde E^\dagger\right)^\#.
\end{eqnarray*}
From this it immediately follows that the condition (\ref{Hinfbound2}) is equivalent to the condition (\ref{Hinfbound1}).  This completes the proof of the lemma. 
\hfill $\Box$
\section{Robust Stability Analysis of an Optical Parametric Amplifier System}
\label{sec:OPA}
\begin{figure}[hbp]
\psfrag{k1}{\color{red} $\kappa_1$}
\psfrag{k2}{\color{blue} $\kappa_2$}
\begin{center}
\includegraphics[width=8cm]{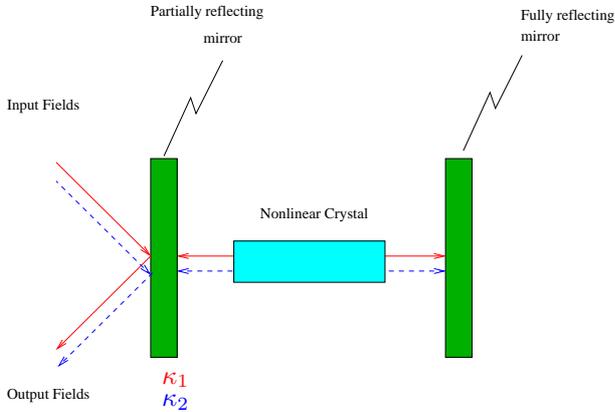}
\end{center}
\caption{Schematic diagram of an OPA system. Here, the red solid lines represent the fields at the fundamental frequency and the blue dashed lines represent the fields at the second harmonic frequency.}
\label{F1}
\end{figure}

In this section, the nonlinear quantum system under consideration is the model of an OPA. An OPA consists of an second-order nonlinear optical medium enclosed in an optical cavity. The second order nonlinear optical medium is referred to as a $\chi^{(2)}$ medium and allows for coupling between a fundamental electromagnetic field and a second harmonic electromagnetic field; e.g., see \cite{BR04,WM08}. The construction of an OPA is illustrated in Figure \ref{F1}. 
A standard $(S,L,H)$ model for an OPA  is as follows:
\begin{eqnarray}
\label{OPA}
S&=&I,~H_1=0,~H_2 = \imath\chi\left(a_2^*a_1^2-\left(a_1^*\right)^2a_2\right),\nonumber \\
~L&=& \left[\begin{array}{c}\sqrt{\kappa_1}a_1\\ \sqrt{\kappa_2}a_2\end{array}\right];
\end{eqnarray}
e.g. see \cite{BR04,WM08}. Here $a_1$ and $a_2$ are the creation operators for the fundamental and second harmonic modes respectively. Also, $\kappa_1 > 0$ and $\kappa_2 > 0$ are parameters defining the reflectivity of the partially reflecting mirror at the fundamental and second harmonic frequencies respectively. Furthermore, $\chi > 0$ is a parameter defining the strength of the $\chi^{(2)}$ nonlinearity. 

The Hamiltonian in this model is a non-quadratic Hamiltonian and so corresponds to nonlinear quantum system. The first term in the expression for $H_2$ can be interpreted as the annihilation of two photons at the fundamental frequency and the creation of a single photon at the second harmonic frequency. Similarly, the second term in the expression for $H_2$ can be interpreted as the annihilation of a single photon at the second harmonic frequency and the creation of two photons at the fundamental frequency. 

We will apply the theory developed in this paper to analyze the robust stability of this nonlinear quantum system. 
We first attempt to apply the results of Theorem \ref{T4} directly to  this quantum system. Hence, we let
\[
M = 0
\]
and
\begin{equation}
\label{OPAf}
f(z,z^*)=\imath\chi\left(z_2\left(z_1^*\right)^2-z_1^2\left(z_2^*\right)\right)
\end{equation}
where $z_1 = a_1^*$ and $z_2 = a_2^*$. 
This defines a nonlinear quantum
system of the form considered in Theorem \ref{T4} with $M_1 = 0$, $M_2 =
0$, $N_1 = \left[\begin{array}{cc}\sqrt{\kappa_1} & 0\\0 & \sqrt{\kappa_2}\end{array}\right]$, $N_2 = 0$, $E_1=0$, $E_2 = I$.  We now investigate whether this function $f(\cdot)$ satisfies the conditions 
(\ref{sector4a}) and  (\ref{sector4b}). First we calculate
\begin{eqnarray*}
\frac{\partial f(z,z^*)}{\partial z_1} &=& -2\imath\chi z_1z_2^*,\quad \frac{\partial^2 f(z,z^*)}{\partial z_1^2} = -2\imath\chi z_2^*,\\
\frac{\partial f(z,z^*)}{\partial z_2} &=& \imath\chi\left(z_1^*\right)^2,\quad \frac{\partial^2 f(z,z^*)}{\partial z_2^2} = 0.
\end{eqnarray*}
From this, we can immediately see that the conditions (\ref{sector4a}) and  (\ref{sector4b}) will not be globally satisfied. 

In order to overcome this difficulty, we first note that any physical realization of a $\chi^{(2)}$ optical  nonlinearity will not be exactly described by the model (\ref{OPA}) but rather will exhibit some saturation of the nonlinear effect. In order to represent this effect, we could assume that the true  function $ f(\cdot)$ describing the Hamiltonian of the OPA  is such that the first two non-zero terms in its Taylor series expansion (\ref{H2nonquad})  correspond to the standard $\chi^{(2)}$ Hamiltonian defined by (\ref{OPAf}). Furthermore, we could assume that the true function $f(\cdot)$ is such that the conditions 
 (\ref{sector4a}) and  (\ref{sector4b}) are  satisfied for suitable values of the constants 
 $\gamma > 0$, $\delta_1\geq 0$, $\delta_2\geq 0$. Here the quantity $\frac{1}{\gamma}$ will be proportional to the saturation limit. 

An alternative approach to dealing with the issue that the conditions (\ref{sector4a}) and  (\ref{sector4b}) will not be globally satisfied by the function $f(\cdot)$ defined in (\ref{OPAf}) is to assume that these conditions only hold over some ``domain of attraction'' and then only conclude robust asymptotic stability within this domain of attraction. This approach requires a semi-classical interpretation of the function $f(\cdot)$ since formally the operators $a_1$ and $a_2$ are unbounded operators. However, it leads to results which are consistent with the known physical behavior of an OPA in that it can become unstable and oscillate if the magnitudes of the driving fields are too large; e.g., see \cite{WM08}. In practice, the true physical situation will combine aspects of both solutions which we have mentioned but we will concentrate on the second approach involving a semi-classical ``domain of attraction''.

\begin{figure}[hbp]
\begin{center}
\psfrag{l1}{$\|z_1\|^2$}
\psfrag{l2}{$\|z_2\|^2$}
\psfrag{d1}{$\frac{\delta_2}{ 4\chi^2}$}
\psfrag{g0}{$\frac{1}{4\gamma^2 \chi^2}$}
\psfrag{g1}{$\frac{1}{\gamma^2 \chi^2}$}
\psfrag{g2}{$\bar \lambda$}
\includegraphics[width=8cm]{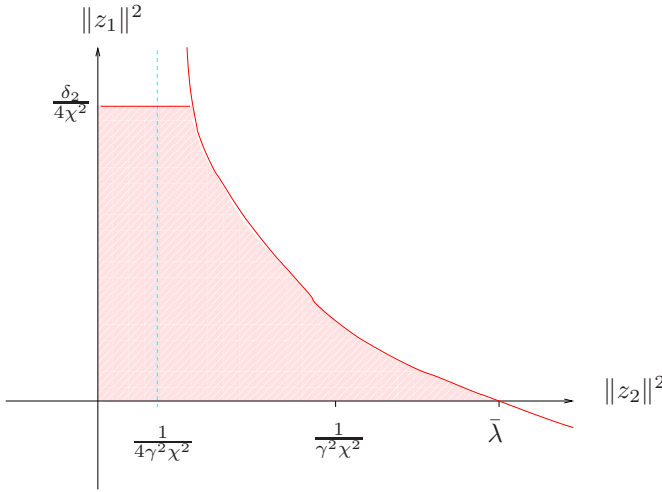}
\end{center}
\caption{Diagrammatic representation of the region for which the constraints (\ref{sector4a}) and  
(\ref{sector4b}) are satisfied. Here $\bar \lambda =\frac{1}{2\gamma^2 \chi^2}+\sqrt{\frac{1}{4\gamma^4 \chi^4}+\delta_1}$. }
\label{F2}
\end{figure}

In order to calculate the region on which our theory can be applied, we note that for our OPA model, 
the condition (\ref{sector4a}) will be satisfied if 
\begin{eqnarray}
\label{domain_1}
4 \chi^2 \|z_1\|^2 \|z_2\|^2 + \chi^2 \|z_1\|^4 
&\leq& \frac{\|z_1\|^2 + \|z_2\|^2}{\gamma^2} + \delta_1;\nonumber \\
\iff \hspace{3cm} \nonumber \\
\|z_2\|^2\left(4  \|z_1\|^2 - \frac{1}{\gamma^2 \chi^2}\right) 
&\leq& \|z_1\|^2 \left(\frac{1}{\gamma^2 \chi^2}- \|z_1\|^2\right)\nonumber \\
&&+\frac{\delta_1}{\chi^2}.
\end{eqnarray}
We now consider the case in which $\|z_1\|^2 >  \frac{1}{4\gamma^2 \chi^2}$. In this case, the condition (\ref{domain_1}) is equivalent to the condition
\begin{eqnarray}
\label{domain_2}
\|z_2\|^2 \leq \frac{\frac{\delta_1}{\chi^2}+\frac{\|z_1\|^2}{\gamma^2 \chi^2}-\|z_1\|^4}{4  \|z_1\|^2 - \frac{1}{\gamma^2 \chi^2}}. 
\end{eqnarray}
In the case that $\|z_1\|^2 \leq \frac{1}{4\gamma^2 \chi^2}$, the left hand side of (\ref{domain_1}) is always negative and the right hand side of (\ref{domain_1}) is always positive. Hence in this case, the condition (\ref{domain_1}) will always be satisfied. Also, the condition (\ref{sector4b}) will be satisfied if 
\begin{equation}
\label{domain_3}
\|z_2\|^2 \leq \frac{\delta_2}{ 4\chi^2}. 
\end{equation}
The conditions (\ref{domain_2}) and (\ref{domain_3}) define the region to which our theory can be applied in guaranteeing the robust mean square stability of the OPA system. This region is represented diagrammatically in Figure \ref{F2}. The constraints (\ref{domain_2}) and (\ref{domain_3}) can be interpreted as bounds on the average values of the internal cavity fields for which robust mean square stability can be guaranteed; see also \cite{SHHPJ2a,SaP2a,SaPH1a}.

We now investigate the strict bounded real conditions (\ref{Hurwitz1}), (\ref{Hinfbound2}). For this system, it follows from the definition (\ref{Hurwitz1}) that the matrix $F$ is given by
\[
F = \left[\begin{array}{cccc}-\frac{\kappa_1}{2} & 0 & 0 & 0\\0 & -\frac{\kappa_2}{2} & 0 & 0\\
0& 0 & -\frac{\kappa_1}{2} & 0\\0 & 0 & 0 & -\frac{\kappa_2}{2}\end{array}\right]
\]
which is Hurwitz for all $\kappa_1 > 0$ and $\kappa_2 > 0$. Thus, the condition (\ref{Hurwitz1}) is always satisfied. Also, we calculate the transfer function matrix $\tilde E\left(sI -F\right)^{-1}J\tilde E^\dagger$ as
\[
\tilde E\left(sI -F\right)^{-1}J\tilde E^\dagger = \left[\begin{array}{cc}
\frac{-1}{s+\frac{\kappa_1}{2}} & 0 \\0 & \frac{-1}{s+\frac{\kappa_2}{2}}
\end{array}\right].
\]
It is straightforward to show that this transfer function matrix has an $H^\infty$ norm of 
\[
\left\|\tilde E\left(sI -F\right)^{-1}J\tilde E^\dagger  \right\|_\infty 
= \sqrt{\max\left\{\frac{4}{\kappa_1^2},\frac{4}{\kappa_2^2}\right\}}.
\]
Thus, for this system, the condition (\ref{Hinfbound2}) is equivalent to the condition 
\begin{equation}
\label{gamma_bound}
\sqrt{\max\left\{\frac{4}{\kappa_1^2},\frac{4}{\kappa_2^2}\right\}} < \frac{\gamma}{2}.
\end{equation}
Hence, using Theorem \ref{T4} and Lemma \ref{L3}, we can conclude that the OPA system (\ref{OPA}) is robustly means square stable provided that the condition (\ref{gamma_bound}) is satisfied and Heisenberg evolution of the quantities $a_1(t)$ and $a_2(t)$ are such that the conditions (\ref{domain_2}) and (\ref{domain_3}) remain satisfied. 

Note that in most experimental situations, $\kappa_2 \geq \kappa_1$; e.g., see \cite{SHHPJ2a}. This means that 
\[
\sqrt{\max\left\{\frac{4}{\kappa_1^2},\frac{4}{\kappa_2^2}\right\}} = \frac{2}{\kappa_1}. 
\]
If we then equate $\frac{2}{\kappa_1} = \frac{\gamma}{2}$, we obtain $\gamma = \frac{4}{\kappa_1}$ which can be substituted into the right hand side of (\ref{domain_2}) to obtain an upper bound on the region for which the conditions (\ref{sector4a}) and  
(\ref{sector4b}) are satisfied. This region is defined by  (\ref{domain_3}) and the inequality
\begin{equation}
\label{domain_4}
\|z_2\|^2 \leq \frac{\frac{\delta_1}{\chi^2}+\frac{\|z_1\|^2\kappa_1^2}{16 \chi^2}-\|z_1\|^4}{4  \|z_1\|^2 
- \frac{\kappa_1^2}{16 \chi^2}}. 
\end{equation}
Also, note that the region defined by (\ref{domain_2}) and (\ref{domain_4}) will only be an upper bound on a domain of attraction for the OPA system. To find an actual domain of attraction for this system, we would need to find an invariant subset contained in the region defined by (\ref{domain_2}) and (\ref{domain_4}). Such an invariant set could be chosen to be an ellipsoidal region defined by the quadratic Lyapunov function arising from the matrix $P$ solving (\ref{QMI2}).

\section{Conclusions}
\label{sec:conclusions}
In this paper, we have extended the robust stability result of \cite{PUJ1a} to the case of non-quadratic perturbations to the Hamiltonian which depend on multiple parameters. This led to a robust stability condition of the form of a multi-variable small gain condition. This condition was then applied  the robust stability
analysis of a nonlinear quantum system consisting of an OPA and the stability region for this system was investigated. 

\end{document}